\documentclass[utf8]{frontiersSCNS} 

\usepackage{url,lineno,microtype}
\usepackage[onehalfspacing]{setspace}
\usepackage{siunitx}
\usepackage{csquotes}
\usepackage{tabu}             
\usepackage{booktabs}  
\usepackage{multirow}
\usepackage{lipsum}
\usepackage{verbatim}

\DeclareSIUnit\pixel{px}
\DeclareSIUnit\male{male}
\DeclareSIUnit\female{female}

\renewcommand{\vec}[1]{\boldsymbol{\mathbf{#1}}}
\newcommand{\mat}[1]{\mathbf{#1}}
\newcommand{\set}[1]{\mathcal{#1}}

\newcommand{\N}{\mbox{\rm \hbox{I\kern-.15em\hbox{N}}}}
\newcommand{\R}{\mathbb{R}}

\renewcommand{\matrix}[1]{\begin{pmatrix} #1 \end{pmatrix}}
\renewcommand{\vector}[1]{\begin{pmatrix} #1 \end{pmatrix}}

\usepackage{draftwatermark}
\SetWatermarkLightness{0.8}
\SetWatermarkText{preprint}

\def\keyFont{\fontsize{8}{11}\helveticabold }
\def\firstAuthorLast{Döllinger and Wolf, {et~al.}} 
\def\Authors{Nina Döllinger\,$^{1,*}$, Erik Wolf\,$^{2,*}$, David Mal\,$^{1,2}$, Stephan Wenninger\,$^{3}$, Mario Botsch\,$^{3}$, Marc Erich Latoschik\,$^{2}$ and Carolin Wienrich\,$^{1}$}

\def\Address{$^{1}$Human-Technology-Systems Group, University of Würzburg, Würzburg, Germany \\
$^{2}$Human-Computer Interaction Group, University of Würzburg, Würzburg, Germany \\
$^{3}$Computer Graphics Group, TU Dortmund University, Dortmund, Germany}

\def\corrAuthor{Nina Döllinger}
\def\corrEmail{nina.doellinger@uni-wuerzburg.de}
\extraAuth{Erik Wolf \\ erik.wolf@uni-wuerzburg.de \\ \vspace{0.4em} Both authors contributed equally to this research.}

\begin{document}
\onecolumn
\firstpage{1}

\title[Resize Me! Embodying Realistically Modulatable Avatars in VR]{Resize Me! Exploring the User Experience of Embodied Realistic Modulatable Avatars for Body Image Intervention in Virtual Reality}

\author[\firstAuthorLast ]{\Authors} 
\address{} 
\correspondance{} 

\maketitle

\begin{abstract}
\section{}
Obesity is a serious disease that can affect both physical and psychological well-being. Due to weight stigmatization, many affected individuals suffer from body image disturbances whereby they perceive their body in a distorted way, evaluate it negatively, or neglect it. Beyond established interventions such as mirror exposure, recent advancements aim to complement body image treatments by the embodiment of visually altered virtual bodies in virtual reality (VR).
We present a high-fidelity prototype of an advanced VR system that allows users to embody a rapidly generated personalized, photorealistic avatar and to realistically modulate its body weight in real-time within a carefully designed virtual environment. In a formative multi-method approach, a total of 12 participants rated the general user experience (UX) of our system during body scan and VR experience using semi-structured qualitative interviews and multiple quantitative UX measures. By using body weight modification tasks, we further compared three different interaction methods for real-time body weight modification and measured our system's impact on the body image relevant measures body awareness and body weight perception. From the feedback received, demonstrating an already solid UX of our overall system and providing constructive input for further improvement, we derived a set of design guidelines to guide future development and evaluation processes of systems supporting body image interventions.

\scriptsize
\keyFont{\section{Keywords:} Virtual reality, avatar embodiment, user experience, body awareness, body weight perception, body weight modification, body image disturbance, eating and body weight disorders} 
\end{abstract}

\section{Introduction}
Obesity is defined as a complex chronic disease characterized by a drastic overweight and an above-average percentage of body fat \citep{who2021obesity}. In recent decades, its prevalence has more than doubled and is expected to rise further \citep{venegas2020obesity}.
For those affected, it imposes a severe physical and psychological burden. Physically, it is associated with an increased risk of several secondary diseases like diabetes, cardiovascular disorders, or cancer, and with increased mortality from infections \citep{stefan2021global}. Psychologically, individuals deal with external or internalized stigmatization that can lead to a disturbed body image \citep{meadows2018studies}. Noninvasive treatments of obesity often consist of a multi-method approach combining lifestyle and weight-loss interventions with cognitive-behavioral therapy \citep{yumuk2015european}. In addition to practicing beneficial diet and exercise habits, the focus lies on examining one's own body and managing disturbances in body image. Disturbances in body image, as part of a variety of body weight- and eating disorders \citep{rosen2001bodyimage, thompsonMINIREVIEWAssessmentBody1998}, are composed of a misperception of body dimensions (body image distortion) and the inability to like, accept, or value one's own body (body image dissatisfaction) \citep{turbyne2021review}. The treatment methods are various, including a confrontation with one's own body using a mirror reflection or video recordings, as well as self-acceptance exercises such as meditation \citep{ziserEffectiveness2018,stewart2004light}.  

In recent years, novel methods complementing the therapy of body image disturbances based on virtual reality (VR) have been explored \citep{ferrer2013virtual,wiederhold2016virtual,riva2019virtual}. These methods often use 3D models of human beings \citep{horne2020review,turbyne2021review}, which can be called avatar as they represent a particular user \citep{bailensonAvatars2004}. VR in general, and the confrontation with avatars in particular, have great potential to influence human perception and behavior \citep{WienrichBehaveFIT}. Especially the feeling of embodiment towards an avatar holds enormous potential for therapeutic purposes \citep{matamalagomezVirtualBodyOwnership2021}. It increases the emotional response to important virtual stimuli \citep{gall2021emotional}, and users tend to adapt their behavior and attitudes to the characteristics of their embodied avatar \citep{ratan2019proteus, yeeProteusEffectEffect2007}. 
In the context of body image, avatars have been utilized to expose users of a VR system to virtual bodies or body parts varying in size and/or shape to investigate the principles of body weight perception \citep{thaler2019role,wolf2020bodyperception,wolf2021embodiment, wolf2022holographic} or to influence the perception or attitude towards the user's own body \citep{turbyne2021review}. \citet{dollinger2019vitras} suggested simulating body weight changes using modulatable, photorealistic, personalized, and embodied avatars, and \citet{turbyne2021review} further stated that the users' body image may conform to these avatars after being embodied to it. Recent developments in computer graphics allow the generation of photorealistic avatars that exactly match a person's real-life appearance within a short duration and at low-cost \citep{achenbach2017fast,wenninger2020smartphone, bartl2021lowcost}. Likewise, methods for a realistic modulation of the generated avatars' body dimensions have been established \citep{piryankova2014BodyVisualizer,hudson2020development,maalin2020beyond}. However, to the best of our knowledge, no work to date combines all the above introduced promising technologies in a joint system intending to support body image therapy.

To address this gap, we developed and evaluated a VR system allowing users to embody their photorealistic, personalized avatar and actively modify their avatar's body weight in real-time. The current work aims to present our system development, evaluate the user experience of this interactive approach, and reveal first insights into its usefulness for measuring body weight perception and its impact on body image and body awareness.  
We used an optimized photogrammetry approach to create avatars that reflect the user's exact physical appearance. We carefully designed our system to ensure accessibility and ease of use for future use cases.
As our technical system was developed to be evaluated with potential patients in a clinically relevant setting as part of a later feasibility study within our research project ViTraS \citep[Virtual Reality Therapy by Stimulation of Modulated Body Perception,][]{dollinger2019vitras}, we aimed to guarantee high usability and user experience for future testing and usage with individuals of the target population. For this purpose, we performed a formative user evaluation with a small sample of participants who did not suffer from a disturbed body image. A total of 12 participants underwent a structured experimental procedure, which included the avatar generation process and exposure to their avatar in a virtual therapy office. In VR, they estimated their modulated avatars' body weight and interactively simulated different weight changes using gestures, a joystick, or virtual buttons. We conducted short interviews before and after the virtual experience to capture qualitative feedback. Additionally, participants quantitatively rated the usability of the calibration process and the interaction methods and reported their feeling of presence, embodiment, body awareness, and general perception of their avatar. Finally, we captured their performance during the interactive estimation task to measure body weight perception. Based on our results, we present a set of design guidelines for the future development of similar avatar-based body image therapy support tools.

\section{Related Work}
Body image disturbance is characterized by an \enquote{excessively negative, distorted, or inaccurate perception of one's own body or parts of it} \citep{who2019classification}. It may manifest into body image distortion, the misperception of one's body weight and dimensions that have repeatedly been reported based on underestimations \citep{maximova2008you,valtolina1998body} or overestimations \citep{thaler2018bse,docteur2010overestimation}, or body image dissatisfaction, a negative attitude towards the own body that is associated with body image avoidance \citep{walker2018meta} and a reduced body awareness \citep[awareness for bodily signals,][]{mehling2011body, peat2011self,todd2019exploration,todd2019multiple,zanetti2013clinical}. While often caused by internalized weight stigma and a fear of being stigmatized by others \citep{meadows2018studies}, a disturbed body image interferes with efforts to stabilize body weight in the long term \citep{rosen2001bodyimage}. 
Treatments for body image disturbances mainly rely on cognitive-behavioral therapy, typically combining psychoeducation and self-monitoring tasks, mirror exposure, or video feedback \citep{farrelEmpiricallyEvaluatedTreatments2006, ziserEffectiveness2018, griffen2018mirror}. 
Based on the fundamentals of these established methods, an increasing number of researchers have started to explore VR applications as additional support for attitude and behavior change in general \citep{WienrichBehaveFIT} and therapy of body image disturbance \citep{rivaThevirtualenvironment1997,ferrer2009validity,ferrer2013virtual,riva2019virtual,turbyne2021review} and obesity in particular \citep{horne2020review,dollinger2019vitras}.

\subsection{The Unique Potential of Modulatable Avatars in VR}
VR offers the opportunity to immerse oneself in an alternative reality and experience scenarios that are otherwise only achievable via imagination. Endowed with this unique power, mainly the use of avatars has attracted attention in treating body image disturbance \citep{turbyne2021review, horne2020review}. 
The use of avatars allows the simulation of rapid changes in body shape or weight, enabling further investigation of body weight perception in general \citep{thaler2019role,wolf2020bodyperception,wolf2021embodiment}. While some researchers are using multiple generic avatars differing in body weight \citep{normand2011multisensory,piryankova2014owning,keizer2016virtual,preston2018implicit,ferrer2018embodiment}, others have developed methods for dynamic body weight modification  \citep{alcaniz2000new,johnstoneAssessmentBodyImage2008,nimcharoen2018,piryankova2014BodyVisualizer,hudson2020development,maalin2020beyond,neyret2020bodyperception}. A huge advantage when using advanced body weight modification methods is that the avatar's body weight can be realistically changed to a desired numeric reference value. For this purpose, mainly the body mass index \citep[calculated as $\textrm{BMI} = \frac{\textrm{Body Weight in \si{kg}}}{(\textrm{Body Height in \si{m}})^2}$,][]{who2000obesity} is used.
One example is the work of \citet{thaler2018bse}, who trained a statistical model to apply realistic BMI-based body weight modification to their generated personalized, photorealistic humans. But also other factors like muscle mass could be included into such models \citep{maalin2020beyond}.  However, to our knowledge, previous works have only presented participants with static modified avatars (usually in A-pose). Our system allows users to dynamically alter their body weight while being embodied in VR. The statistical models of weight gain/loss are usually trained on the whole body \citep{piryankova2014BodyVisualizer} or neglect the head region completely \citep{maalin2020beyond}. We also learn a statistical model of weight gain/loss, but keep a small part of the face region fixed in order to better preserve the identity of the users when applying the body weight modification.

Besides the shape of the used avatar, application or system-related properties also might alter how we perceive body weight in VR. \citet{wolf2020bodyperception} presented an overview of potentially influencing factors. While factors like the used display or the observation perspective might unintentionally alter body weight perception, especially the personalization and embodiment of avatars hold potential for application in body image interventions. For example, by using realistic and modulatable avatars, \citet{thaler2018bse} investigated the influence of personalization on the perception of body weight. They found that the estimator's BMI influences body weight estimations of an avatar, but only when the avatar's shape and texture matched the estimator's appearance. This comes along with a recent review of \citet{horne2020review}, who identified the personalization of avatars as an important factor when using avatars.

In addition to avatar personalization, the feeling of embodiment towards one's avatar is particularly interesting for avatar-based body weight estimation scenarios. It can be evoked by visuomotor coherence, for example, when the virtual body moves like the real body \citep{slaterInducingIllusoryOwnership2009,slater2010first} and is divided into the feeling of being inside (self-location), of owning (virtual body ownership), and of controlling (agency) an avatar \citep{kilteni2012embodimentinvr}. However, the feeling of embodiment interferes with the user's physical body awareness. \citet{tsakiris2011just} showed that individuals with a low body awareness are likely to perceive a higher feeling of embodiment towards an avatar than individuals with high body awareness. \citet{filippetti2017heartfelt} showed that individuals with low body awareness experience increased body awareness when embodying a virtual avatar. In a recent experiment, \citet{doellinger2022bodymind} revealed that especially the feeling of body ownership towards a virtual body was positively related to an increase in body awareness. Thus, embodiment offers the possibility of strengthening body awareness as a potential mediator for a positive body image. However, there is no research on the connection between embodying avatars and body awareness in a VR-based body image treatment task.

In terms of body weight perception, \citet{wolf2021embodiment} recently found that a female's own BMI influences body weight estimations of a generic avatar only when embodying it. \citet{jung2020time} showed that changes in the body weight of a self-embodied avatar are less likely to be noticed. \citet{riva2019virtual} further stated that embodiment could potentially help to update the wrong representation of one's physical body by experiencing the ownership over a virtual body with a different shape or size. This statement goes along with the work of \citet{scarpina2019effect}, who investigated the influence of being embodied in a thinner generic avatar. They showed that both normal weight and obese users estimated the circumference of their real hips lower after exposition to the embodied avatar, but not without being embodied. \citet{turbyne2021review} highlighted that participants’ body image conformed to modified virtual body size when participants felt embodied in it. This mechanism can be attributed to the Proteus effect \citep{yeeProteusEffectEffect2007, ratan2019proteus}, suggesting that people tend to conform to their embodied avatars' characteristics both in behavior and attitudes.

\subsection{User Interaction for Body Weight Modification}
Most VR applications for body image interventions aim for enhanced mirror confrontation. They surpass real mirror confrontation by modifying the mirror image or the shown avatars into different body shapes. In our work, we want to go one step further and allow users to adjust the shape of their avatar interactively. Our idea is to give the users the opportunity to actively engage in analyzing their body image and develop a novel feeling for their own body. Object manipulation in VR has been widely researched and can serve as a reference in the development of body weight modification interaction methods. For example, \citet{laviola20173d} presented a set of design guidelines for different types of object manipulation, including object scaling by virtual buttons or other control elements, the inclusion of physical interfaces as provided on most VR controllers, or the design of gesture-based object manipulation. Furthermore, \citet{williams2020gestures} and \citet{wu2019understanding} investigated the preference of users towards different gestures in object manipulation, and both proposed using two-handed gestures (e.g., moving the hands apart or bringing them together) for size modification of large objects. These results could be applied to the manipulation of body weight in VR. Another option, which could be especially beneficial for novice users due to its high naturalness and intuitiveness, are multimodal interfaces \citep{wolf2019multimodal,zimmerer2020multimodal}. However, due to the still low complexity of our interface, we have refrained from integrating them into our system for now. In summary, and to the best of our knowledge, there has been no specific research on interactive real-time body weight modification so far.

\subsection{Summary and Outline}
When considering the introduced research, it suggests to further work on complementing established treatment methods for body image disturbances with the benefits of avatar interaction. 
Former research mainly focused on the general perception of body weight in VR or the impact of differently shaped avatars on the perception of the own body. However, while highly developed avatar weight modification algorithms exist to date, only few systems combine the potentials of simulated body weight modifications with embodiment or user interaction. Additionally, most current investigations have not considered body awareness as a possible mediator between the feeling of embodiment and body image. Further, it has not yet been investigated whether an interactive avatar modification approach has an additional impact on body image.
Our current work within the interdisciplinary research project ViTraS \citep{dollinger2019vitras} addresses these research gaps and aims towards a novel approach for supporting body image therapy using a user-centered design process.
In the current work, we present a high-fidelity prototype of a body image therapy support system that allows users to embody their rapidly generated personalized, photorealistic avatar within a carefully designed virtual environment. Users can realistically modify their avatars’ body weight using three different interaction methods (joystick, gestures, and virtual objects). We focus on a user experience evaluation with normal-weight participants performed within our first system design cycle. In a comprehensive mixed-methods evaluation, we assessed (1) the body scan experience during the avatar generation process, (2) the user experience (UX) of the VR exposure and different modification methods, and (3) their impact on body image-related outcomes, including body awareness and body weight misestimation. To sum up our results, we derive a set of guidelines for the design and implementation of future VR systems to support body image interventions.
\section{System Description}
The technical implementation of our system is realized using the game engine Unity version 2019.4.15f1~LTS~\citep{unity2019gameengine}. As VR HMD, we use a Valve Index \citep{valve2020index}, providing the user a resolution of 1440$\times$1600\,pixels per eye with a total field of view of \SI{120}{\degree} running at a refresh rate of \SI{90}{\Hz}. For motion tracking, we use the two handheld Valve Index controllers, one HTC Vive Tracker 3.0 positioned on a belt at the lower spine, and two HTC Vive Tracker 3.0 on each foot fixed by a velcro strap. All VR hardware is integrated using SteamVR in version~1.16.10 \citep{value2020steamvr} and its corresponding Unity plugin in version~2.7.3\,\footnote{\url{https://assetstore.unity.com/packages/tools/integration/steamvr-plugin-32647}}. In our evaluation, the system was driven by a high-end PC composed of an Intel Core i7-9700K, an Nvidia RTX2080 Super, and 32\,GB RAM running Windows 10. To ensure that users always received a fluent VR experience and to preclude a possible cause of simulator sickness, we measured the motion-to-photon latency of our system by frame-counting \citep{he2000framecounting}. For this purpose, the video output was split into two signals using an Aten VanCryst VS192 display port splitter. One signal still led to the HMD, the other to an ASUS ROG SWIFT PG43UQ low-latency gaming monitor. A high-speed camera of an iPhone 8 recorded the user's motions and the corresponding reactions on the monitor screen at 240\,fps. The motion-to-photon latency for the HMD matched the refresh rate of the used Valve Index closely, as it averaged \SI{14.4}{\ms} ($SD=\SI{2.8}{\ms}$). The motion-to-photon latency for the body movements was considered low enough to provide a high feeling of agency towards the avatar \citep{waltemate2016latency}, as it averaged \SI{40.9}{\ms} ($SD=\SI{5.4}{\ms}$). A video showing the application is provided in the supplementary material.

\subsection{Virtual Environments}
\begin{figure}[!htb]
 \centering
 \includegraphics[width=\textwidth,trim={0mm 20mm 0mm 40mm},clip]{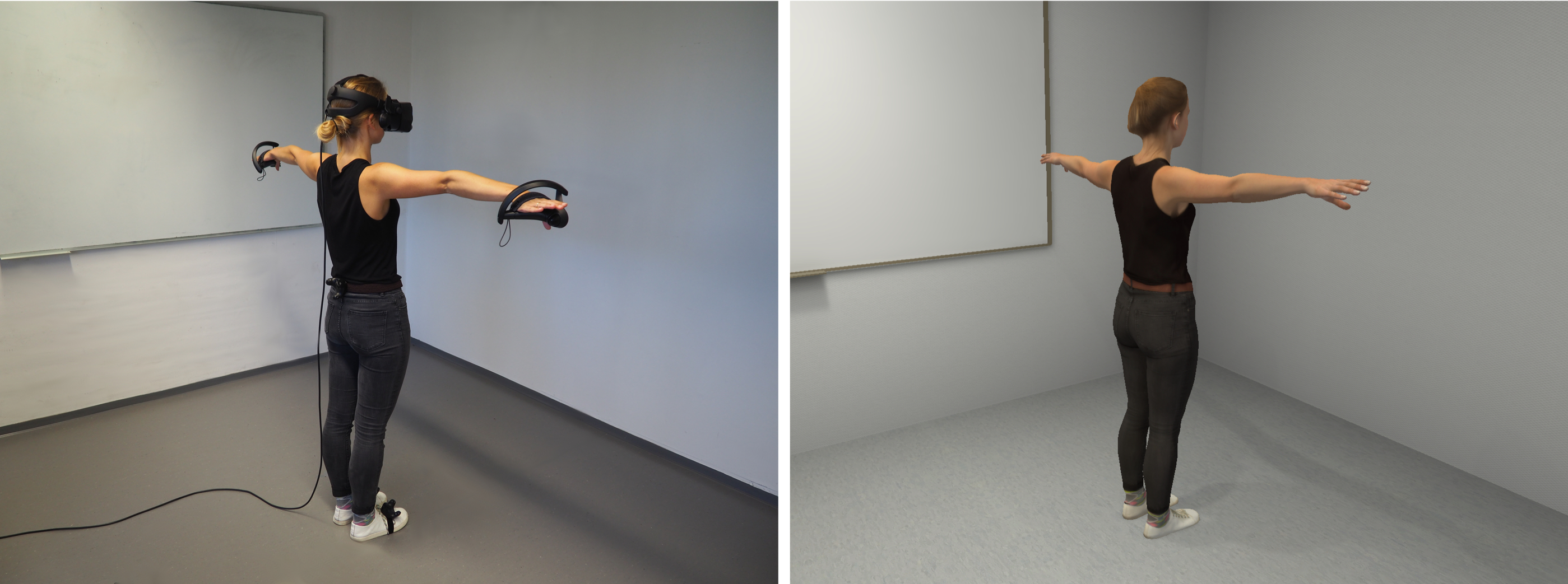}
 \caption{The figure depicts a comparison between the real environment where the experiment took place (left) and the replicated virtual environment used for preparation (right). Both environments contain a user, respectively the avatar, performing the embodiment calibration.}
 \label{fig:e31}
\end{figure}

We realized two virtual environments. The first environment replicates the real environment, in which the user was located physically during our evaluation, and is automatically calibrated accurately to overlay the physical environment spatially (see Figure~\ref{fig:e31}). Here, all preparatory steps required for exposure are performed and tested (e.g., ground calibration, vision test, equipment adjustments, embodiment calibration).
For spatial calibration, we use a customized implementation of the Kabsch algorithm\,\footnote{\url{https://github.com/zalo/MathUtilities\#kabsch}}  \citep{mueller2016kabsch}, based on the positions of the SteamVR base stations in real and virtual environments. Additionally, the virtual ground height is calibrated by briefly placing the controller onto the physical ground.

\begin{figure}[!htb]
 \centering
  \includegraphics[width=\textwidth]{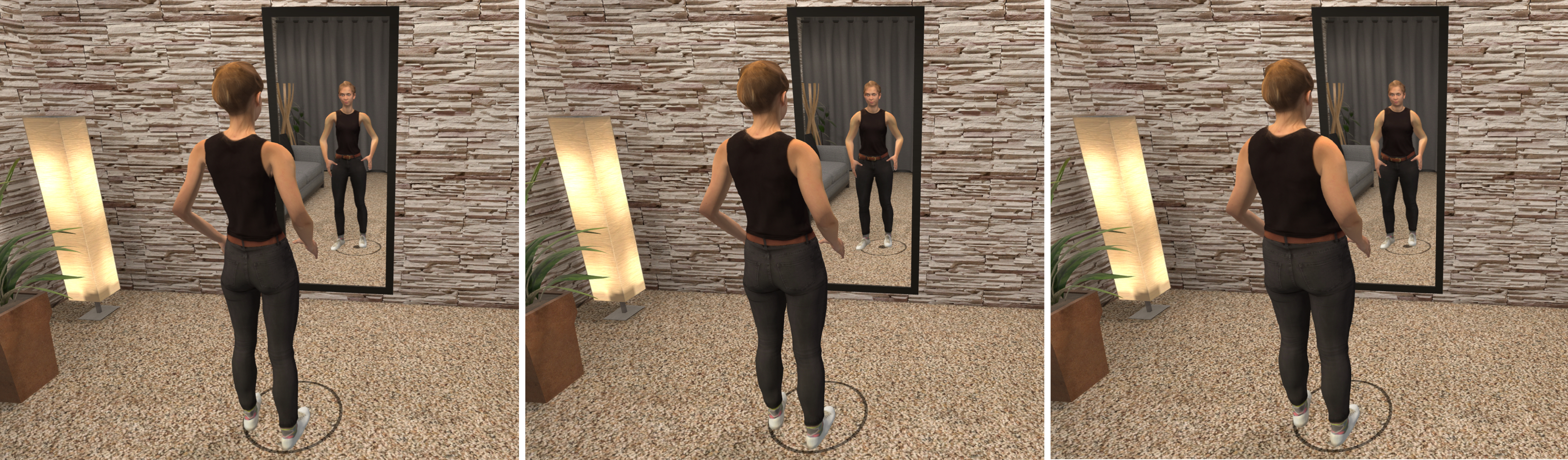}
  \caption{The images show a participant's personalized avatar standing in front of a mirror within the virtual exposition environment of our concept prototype with a reduced (left), normal (center), or increased (right) body weight.}
 \label{fig:teaser}
\end{figure}

The second environment is originally based on an asset taken from the Unity asset store\,\footnote{\url{https://assetstore.unity.com/packages/3d/props/interior/manager-office-interior-107709}} that was modified to match our requirements. This exposition environment is inspired by a typical office of a psychotherapist with a desk and chairs and an exposure area in which the mirror exposure takes place (see Figure \ref{fig:teaser}). The exposure area includes a virtual mirror allowing for an allocentric observation of the embodied avatar. We aimed for a realistic and coherent virtual environment to enhance the overall plausibility of the exposure  \citep{slater2009place,latoschik2021plausibility}.

\subsection{Generation and Animation of Personalized Avatars}
In our system, the user embodies a personalized avatar from an egocentric perspective while the avatar is animated according to the user's body movements in real-time. The following sections describe the generation of the avatars as well as the animation system.

\subsubsection{Generation Process}
\label{sec:bodyscan}
The generation of the avatars closely follows the method of \citet{achenbach2017fast}. First, the subject is scanned with a custom-made photogrammetry rig consisting of 94 DSLR cameras, where four studio lights equipped with diffuser balls ensure uniform illumination \citep[see][Figure 1, top]{bartl2021lowcost}. Instead of employing a separate face scanner like \citet{achenbach2017fast} did, 10 of the 94 DSLR cameras are zoomed in on the subject's face to capture more detail in this region. The images taken by the cameras are then automatically processed with the commercial photogrammetry software Agisoft Metashape \citep{metashape}, resulting in a dense point cloud of the subject. We manually select 23 landmarks on the point cloud in order to guide the subsequent template fitting process. The counterparts of these landmarks are pre-selected on the template model, which comes from the Autodesk Character Generator~\citep{autodesk_chargen} and consists of $N \approx 21k$ vertices, an animation skeleton with skinning weights, facial blendshapes, as well as auxiliary meshes for eyes and teeth. \citet{achenbach2017fast} enhance the template with a statistical model of human shape variation. This statistical, animatable human template model is then fitted to the point cloud by optimizing for alignment, pose, and shape by employing non-rigid ICP \citep{registration-tutorial}. This optimization of the model parameters defines the initial registration of the template, which is then further refined by allowing fine-scale deformation of the vertices to match the scanner data more closely. For more details, we refer to \citet{achenbach2017fast}.

\subsubsection{Animation Process}
For avatar animation, the participants' movements are continuously captured using the SteamVR motion tracking devices. The tracking solution provides for our work a sufficiently solid and rapid infrared-based tracking solution for the crucial body parts required for animation without aligning different tracking spaces \citep{niehorster2017accuracy}. To calibrate the tracking devices to the user’s associated body parts and capture the user’s body height, arm length, and current limb orientations, we use a custom-written calibration script that requires the user to stand in T-pose for a short moment (see Figure \ref{fig:e31}). The calibrated tracking targets of the head, left hand, right hand, pelvis, left foot, and right foot were then used to drive an inverse kinematics \citep[IK,][]{aristidou2018inverse} approach realized by the Unity plugin FinalIK version~2.0\,\footnote{\url{https://assetstore.unity.com/packages/tools/animation/final-ik-14290}}. FinalIK's integrated VRIK solver continuously calculates the user's body pose according to the provided tracking targets. The tracking pose is automatically adjusted to the determined body height and arm length in order to match the user's body. 
In the next step, the tracking pose is continuously retargeted to the imported personalized avatar. Potentially occurring inaccuracy in the alignment of the pose or the end-effectors can be compensated by a post-retargeting IK-supported pose optimization step. This leads to high positional conformity between the participant's body and the embodied avatar and avoids sliding feet due to the retargeting process. 

\subsection{Body Weight Modification of Avatars}
Our system allows the user to modify their body weight during runtime dynamically. The statistical model of weight gain/loss and the implemented user interaction methods are described in the following.

\subsubsection{Data-driven Body Weight Modification}
\label{sec:body-weight-modification}
To build a statistical model of body weight modification, we follow the approach of \citet{piryankova2014BodyVisualizer}, who first create a statistical model of body shape using Principal Component Analysis (PCA) and then estimate a linear function from anthropometric measurements to PCA coefficients. For computing the statistical model of human body shape, we use the template fitting process described above to fit our template model to the European subset of the CAESAR scan database  \citep{robinette2002caesar}. It consists of $M = 1700$ 3D scans, each annotated with anthropometric measurements such as weight, height, arm span, inseam, waist width, etc. After bringing the scans into dense correspondence via template fitting, we are left with $M$ pose-normalized meshes consisting of $N$ vertices each. Our approach for data-driven weight gain/loss simulation differs from the method of \citet{piryankova2014BodyVisualizer} in the following ways: (1) instead of encoding body shape as a $3 \times 3$ deformation matrix per mesh face \citep{Anguelov2015Scape}, we encode body shape directly via vertex positions, (2) we keep vertices in the face region fixed while deforming the rest of the mesh in order to better preserve the identity of the participants.

\begin{figure}[!htb]
    \centering
    \includegraphics[width=\textwidth]{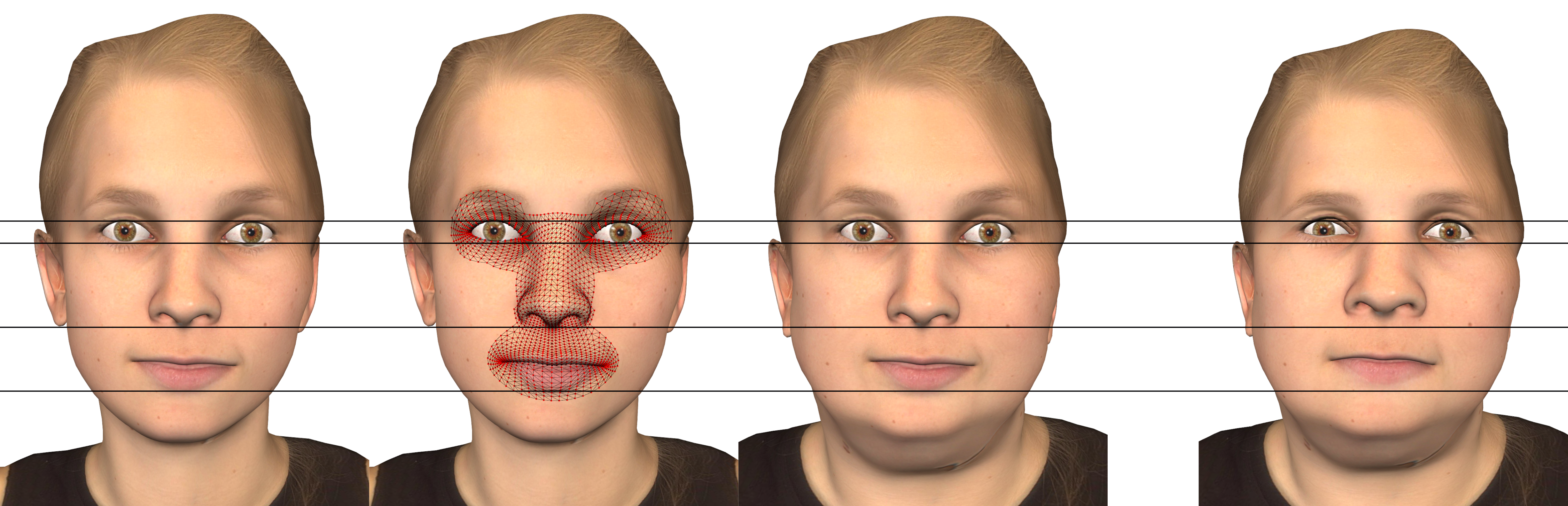}
    \caption{The figure illustrates our approach of facial weight gain simulation. When modifying the weight of an avatar (left), part of the face region (highlighted in red) is fixed (center left). The modified vertices are stitched to the face region in a seamless manner using differential coordinates \citep{Sorkine2015LaplacianMeshProcessing} (center right). Not keeping these vertices fixed would require recalculating the position of all auxiliary meshes such as eyes and teeth due to the undesired change in shape for nose, mouth and eyes (right). For the right image, eyes are copied from the unmodified avatar in order to better highlight the change in shape and position.}
    \label{fig:FixedFaceRegion}
\end{figure}

To this end, we define a set $\set{S}$ with cardinality $V$ containing all vertices outside of the face region (see Figure \ref{fig:FixedFaceRegion}) as well as a selector matrix $\mat{S} \in \R^{3V \times 3N}$ which extracts all coordinates belonging to vertices in $\set{S}$. Let $\vec{x}_j = \left(\vec{p}_1^T, \dots, \vec{p}_N^T \right)^T \in \R^{3N}$ be the vector containing the stacked vertex positions of the $j^\mathrm{th}$ training mesh and $\bar{\vec{x}} = \frac{1}{M} \sum_j \vec{x}_j \in \R^{3N}$ be the corresponding mean. Performing PCA on the data matrix $\mat{X} = \matrix{\mat{S}\vec{x}_1 - \mat{S}\bar{\vec{x}}, \dots, \mat{S}\vec{x}_M - \mat{S}\bar{\vec{x}}} \in \R^{3V \times M}$ and taking the first $k = 30$ components then yields the PCA matrix $\mat{P} \in \R^{3V \times k}$. Let $\mat{W} = \matrix{\vec{w}_1, \dots, \vec{w}_M} \in \R^{k \times M}$ contain the PCA coefficients $\vec{w}_j$ of the $M$ training meshes, computed by $\vec{w}_j = \mat{P}^T \left(\mat{S}\vec{x}_j - \mat{S}\bar{\vec{x}} \right)$. If we denote by $\mat{D} \in \R^{M \times 4}$ the matrix containing the anthropometric measurements weight, height, armspan and inseam of the $j^\mathrm{th}$ subject in its $j^\mathrm{th}$ row, we can then compute a linear mapping from anthropometric measurements $\mat{D}$ to PCA coefficients $\mat{W}$ by solving the linear system of equations $\matrix{\mat{D} \ | \ \vec{1}}\mat{C} = \mat{W}^T$ in a least-squares sense via normal equations.

\begin{figure}[!t]
 \centering
 \includegraphics[width=\textwidth,trim={0mm 0mm 0mm 0mm},clip]{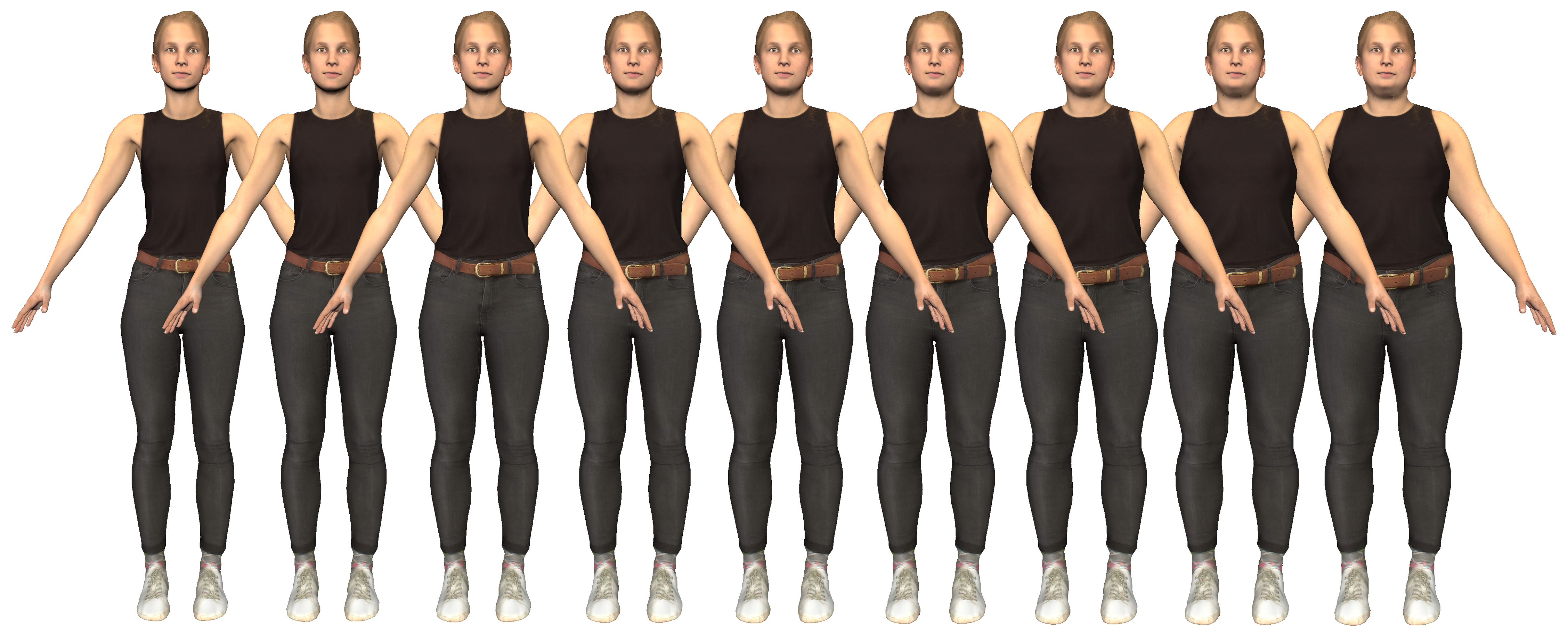}
 \caption{The figure shows a generated female avatar ($\textrm{BMI} = 19.8$) with modified body weight corresponding to a BMI range of 16 to 32 in two-point increments.}
 \label{fig:therapyroom-avatar-comparison}
\end{figure}

New vertex positions for a subject with initial vertex positions $\vec{x}$ and a desired change in anthropometric measurements $\Delta\vec{d} \in \R^{5}$ can then be calculated via $\tilde{\vec{x}} = \mat{S}\vec{x} + \mat{P}\left(\mat{C}^T\Delta\vec{d}\right)$, i.e., by first projecting the desired change in measurements into PCA space via the learned linear function and then into vertex position space via the PCA matrix. However, this only updates vertices in $\set{S}$. In order to seamlessly stitch the new vertex positions to the unmodified face region, we compute the Laplacian coordinates (discretized through cotangent weights and Voronoi areas \citep{Botsch10PMP} of the resulting mesh and then use surface reconstruction from differential coordinates \citep{Sorkine2015LaplacianMeshProcessing}. For the vertices of the face region and its 1-ring neighborhood, the Laplacian is computed based on the unmodified vertex positions $\vec{x}$, while for the rest of the vertices, the Laplacian is computed based on the modified vertex positions $\tilde{\vec{x}}$. Since the position of vertices of the face region is known and should not change, we treat the position of these vertices as \emph{hard} instead of \emph{soft} constraints as discussed by \citet{BotschSorkine2008}. Setting $\Delta\vec{d} = \vector{\Delta w, 0,0,0,0}^T$ then allows modifying the user's weight while keeping the other anthropometric measurements fixed. Keeping the face region fixed (1) better preserves the identity of the user for high values of $\Delta w$ and (2) avoids having to recalculate the position of auxiliary meshes of the avatar such as eyes and teeth (Figure \ref{fig:FixedFaceRegion}). Results of the described body weight modification method are shown in Figure \ref{fig:therapyroom-avatar-comparison}.


\subsubsection{Interaction Methods}
\label{sec:interaction-methods}
\begin{figure}[!b]
 \centering
 \includegraphics[width=\textwidth,trim={20mm 10mm 10mm 10mm},clip]{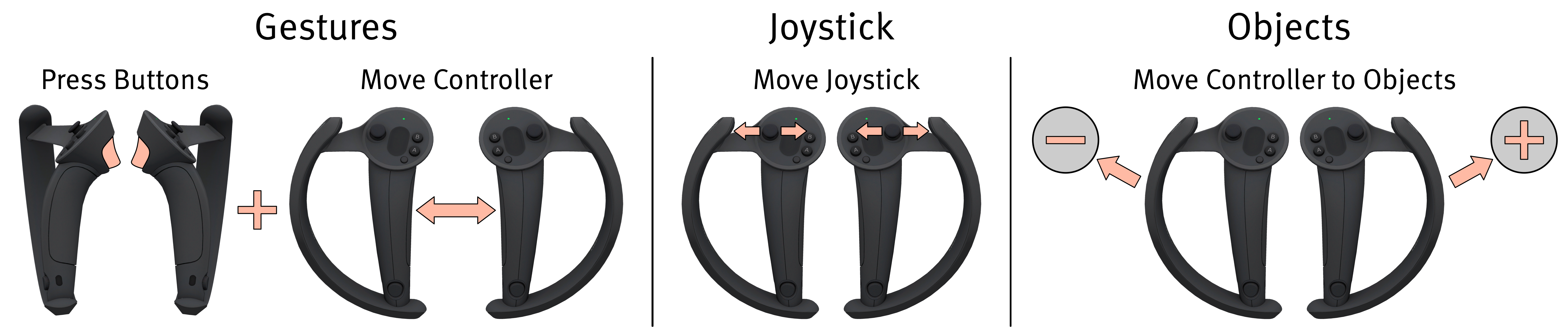}
 \caption{The figure sketches the three body weight modification methods we used for our evaluation: Gestures (left), Joystick (center), and Objects (right).}
 \label{fig:interaction-methods}
\end{figure}

To allow users to modify the avatars' body weight as quickly, easily, and precisely as possible, we compare in our evaluation three implemented interaction methods regarding their usability. Since interaction methods for human body weight modification have not yet been explored, we considered the guidelines for object modification presented by \citet{laviola20173d}. Figure \ref{fig:interaction-methods} gives a short overview of the body weight modification methods. We restricted the body weight modification for all interaction methods to a range of $\pm 35\%$ of the user's body weight to avoid unrealistic, possibly unsettling shape deformation. The constants given in the formulas for calculating the velocity of body weight change were determined empirically in a pilot test.

\paragraph{Body Weight Modification via Controller Movement Gestures}
To modify the avatar's body weight via gestures (see Figure \ref{fig:interaction-methods}, left), users have to press the trigger button on each controller simultaneously. Moving the controllers away from each other then increases the body weight, while moving them towards each other decreases it. The faster the controllers are moved, the faster the body weight changes. When active, the body weight changes by the velocity $v$ in $\si{kg}/\si{s}$, determined by the relative distance change between the controllers $r$ in $\si{m}/\si{s}$, and calculated as $v=3.5r^2+15r$.

\paragraph{Body Weight Modification via Joystick Movement}
To modify the avatar's body weight via joystick (see Figure \ref{fig:interaction-methods}, center), users have to tilt the joystick of either the left or the right controller. Selecting joystick for an initial modification leads to a deactivation of the other joystick for the remaining interaction. Tilting the joystick to the left decreases the body weight, tilting it to the right increases it. The stronger the joystick is tilted, the faster the body weight changes. When enabled, the body weight changes by the velocity $v$ in $\si{kg}/\si{s}$, determined by the normalized tilt $t$ of the joystick and calculated as $v=10t^2+5$. 

\paragraph{Body Weight Modification via Controller Movement Towards Objects}
To modify the avatar's body weight via objects (see Figure \ref{fig:interaction-methods}, right), users have to touch either a virtual \enquote{plus} or a virtual \enquote{minus} object within the virtual environment. As long as an object is touched, the body weight increases or decreases. The longer the object is touched, the faster the body weight changes. When active, the body weight modification velocity $v$ in $\si{kg}/\si{s}$ increases quadratically over a normalized contact duration $d$ of $1.5\,\si{s}$ in a normalized range $r$ between $3\,\si{kg}/\si{s}$ and $15\,\si{kg}/\si{s}$. 
\section{Evaluation}
We tested our first system prototype in a structured UX evaluation based on multiple relevant qualitative questions and quantitative measures concerning the users' scan and VR exposure experience as well as their body image. The following sections contain a detailed explanation of the evaluation process.

\subsection{Ethics}
Since our technical system was developed with the aim of being tested on potential patients in a clinically relevant context as part of a later feasibility study, particular attention has already been paid to ethical aspects during the here reported development and evaluation of our system. As part of a conservative development and evaluation strategy, we decided to work with a relatively small sample of healthy participants in this initial formative evaluation. The system, as well as the evaluation, was designed in consultation with our clinical partners within the context of our interdisciplinary research project ViTraS \citep{dollinger2019vitras}. A detailed ethics proposal following the Declaration of Helsinki was submitted to the ethics committee of the Human-Computer-Media Institute of the University of Würzburg and found to be ethically unobjectionable. Free professional help services provided by the Anorexia Nervosa and Associated Disorders (ANAD)\,\footnote{https://www.anad.de/} organization were explicitly highlighted during the acquisition and evaluation process.

\subsection{Participants}
A total of 12 students (8 female, 4 male) of the University of Würzburg participated in our evaluation and received course credit in return. Before the evaluation, we defined four exclusion criteria queried by self-disclosure: Participants had to have (1) normal or corrected to normal vision and hearing, (2) at least ten years of experience with the German language, (3) not suffered from any kind of mental or psychosomatic disease, or from body weight disorders, and (4) no known sensitivity to simulator sickness. None of the participants matched any exclusion criteria. The participants were aged between 20 and 25 ($M=22.0$, $SD=1.48$), had a BMI between 17.85 and 32.85 ($M=22.72$, $SD=3.98$), and had mostly very little VR experience. Nine out of the twelve participants claimed to know their current body weight. The mean deviation of the indication of their body weight compared to that measured by the experimenter was \SI{0.29}{\kg} ($SD=1.57$).

\subsection{Design}
The evaluation of our system included qualitative and quantitative measures regarding (1) the body scan experience, (2) the UX of the VR exposure and the different modification methods used, and (3) their impact on the body image-related measures body awareness and body weight estimation. 
To compare our three modification methods (see Figure \ref{fig:interaction-methods}), participants performed for each modification method a set of active modification tasks (AMTs) in a counterbalanced order using a 1x3 within-subjects design. For comparing the novel AMT with the more traditional passive estimation task (PET), the participants performed a PET each before and after the AMTs (see Figure~\ref{fig:procedure}, right). All tasks and the timing of the measures will be further explained in Section \ref{sec:procedure}.


\subsection{Measures}
\label{sec:measures}

\subsubsection{Body Scan Experience}
We conducted a semi-structured interview to evaluate the body scan experience. It included questions concerning the participants' expectations, their physical and psychological comfort and/or discomfort during the body scan and the assessment of their body measures, and about the clarity and transparency of the process. A full version of the questions can be found in the supplementary material of this work.

\subsubsection{VR Experience}
Regarding the VR experience, we included a variety of VR-specific and task-specific UX measures to get a holistic view of the system's overall UX, as recommended by previous work \citep{tcha2016questionnaire,wienrich2020appraisevr}. Following \citet{wienrich2020appraisevr}, we used a combination of qualitative and quantitative measures, in virtuo ratings, and pre- and post-questionnaires for the VR UX evaluation.

\paragraph{Interview}
We conducted a semi-structured interview with focus on the VR experience. It included questions concerning the participants' expectations and feelings towards the avatar, their favored body weight modification method and the perceived difficulty of the body weight estimation in general, their intensity of body awareness, and their affect towards their body. A full version of the questions can be found in the supplementary material of this work.

\paragraph{Presence}
We measured the participants' feeling of presence using the Igroup Presence Questionaire \citep[IPQ,][]{schubert2001experience}. It captures presence through 14 questions, each rated on a scale from 0 to 6 (\textit{6 = highest presence}). The items are divided into four different dimensions: general presence, spatial presence, involvement, and realism. The questionnaire was provided directly after the VR exposure to capture presence as accurately as possible.

\paragraph{Embodiment}
As suggested by prior work, we divided the measurements for the feeling of embodiment into VBO and agency \citep{kilteni2012embodimentinvr}. Following \citet{waltemate2018impact} and \citet{kalckert2012moving}, we presented one embodiment question for each dimension on a scale from 0 to 10 (\textit{10 = highest}). To investigate possible differences in the feeling of embodiment caused by our interaction methods, the questions were provided multiple times during exposure.

\paragraph{Simulator Sickness}
To test our system prototype regarding simulator sickness, we included the Simulator Sickness Questionnaire \citep[SSQ,][]{kennedy1993simulator,bimberg2020usage} before and after the VR exposure. It captures the appearance and intensity of 32 different simulator sickness associated symptoms on 4-point Likert scales. The total score of the questionnaire ranges from 0 to 235.62 (\textit{235.62 = strongest}). An increase in the score by 20 between a pre- and post-measurement indicates the occurrence of simulator sickness \citep{stanney1997cybersickness}.

\paragraph{Avatar Perception}
For measuring the affect towards the avatar, we used the revised version of the Uncanny Valley Index  \citep[UVI,][]{hoMeasuringUncannyValley2017}, including its four sub-dimensions: humanness, eeriness, spine-tingling, and attractiveness. Each dimension is captured by four to five items ranging from 1 to 7 (\textit{7 = highest})


\paragraph{Workload}
We measured workload to (1) determine the perceived effort during the calibration of the system and to (2) determine the perceived difficulty when modifying the avatar's body weight with our modification methods. To capture workload fast and efficiently during VR, we used a single item scale ranging from 0 to 220 (220~=~highest) called SEA scale \citep{eilers1986workload}, a German version of the Rating Scale Mental Effort \citep{zijlstra1993efficiency, arnold1999mental}. 

\paragraph{Preference Rankings}
Participants were asked to order the three body weight modification methods concerning their workload, perceived body weight estimation difficulty, vividness, contentment, and overall preference. Ranking scores were then calculated using weighted scores with reversed weights. A weighting of 4 was used for the highest rank, a weighting of 3 for the second highest, and so on. The overall rankings were summed up and averaged over the number of ratings. A high scores states high workload, difficulty, vividness, contentment, and overall preference.

\paragraph{Calibration and Modification Time}
To measure the efficiency of the avatar calibration and the interactions methods, we captured the average time needed from the beginning of calibration or modification until the end. Calibration time included potential amendments of the avatar skeleton and re-calibrations. A lower time states a higher efficiency.

\subsubsection{Body Image}
\paragraph{Body Awareness}
Similar to VBO, agency, and workload, we included (virtual) body awareness (VBA) as a one-item scale from 0 to 10 (\textit{10 = highest VBA}) assessed at multiple times during exposure. The item was derived from the State Mindfulness Scale \citep[SMS,][]{tanay2013state}. 

\paragraph{Passive Body Weight Estimation (PET)}
The PET was adapted from prior work \citep{wolf2020bodyperception,wolf2021embodiment,wolf2022holographic} and used to capture the participants' ability to numerically estimate the avatars' body weight based on a provided body shape. We repeatedly modified the body weight of the embodied avatar within a range of $\pm 20\%$ incremented in $5\%$ intervals in a counterbalanced manner resulting in $n=9$ modifications. To not provide any hints on the modification direction, the HMD was blacked-out during the modification. For each modification, the participants had to estimate the avatar's body weight in $\si{kg}$, which we used to calculate the misestimation $M$. It is based on the estimated body weight $e$ and the presented body weight of the avatar $p$ as $M=\frac{e-p}{p}$. A negative value states an underestimation, a positive value an overestimation. Additionally, we calculated (1) the average misestimation $\overline{M} = \frac{1}{n}\sum_{k=1}^{n}M_k$ and (2) the absolute average percentage of misestimation as $\overline{A} = \frac{1}{n}\sum_{k=1}^{n}|M_k|$.

\paragraph{Active Body Weight Estimation (AMT)}
The AMT was inspired by related work \citep{piryankova2014BodyVisualizer,thaler2018bse,thaler2018visual} and used to capture the participants' ability to modify the avatar's body weight to match a requested numeric value. We also used it to analyze whether a certain interaction method for body weight modification influenced the participant's ability to judge the avatars' body weight. Participants had to modify the avatar's body weight by using one of our modification interaction methods until they thought it matched a presented numeric target weight in $\si{kg}$. The task was repeated for a target weight range of $\pm 20\%$ of the actual avatar's body weight incremented in $5\%$ intervals in a counterbalanced manner resulting in $n=9$ modifications. For each modification, we calculated the misestimation $M$ based on the modified body weight $m$ and the target body weight $t$ as $M = \frac{t-m}{t}$. A negative value states an underestimation, a positive value an overestimation. Additionally, we calculated $\overline{M}$ and $\overline{A}$ as for the PET.



\subsection{Procedure}
\label{sec:procedure}
\begin{figure}[!htb]
 \centering
 \includegraphics[width=0.5\textwidth,trim={0mm 0mm 0mm 0mm},clip]{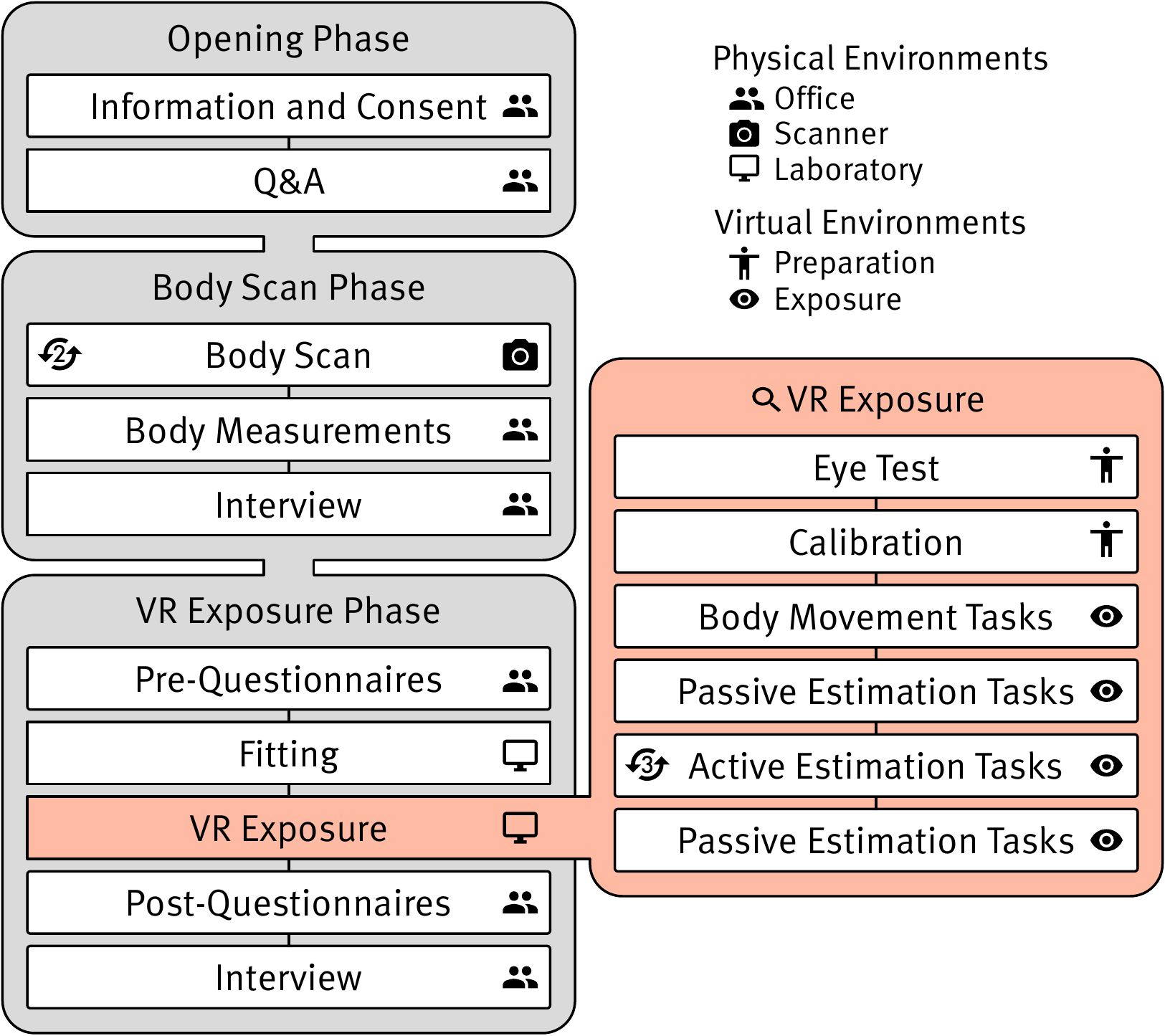}
 \caption{The figure provides an overview of the evaluation process as whole (left) and a detailed overview of the VR exposure (right). The icons on the right side of each step show in which physical or virtual environment the step was conducted. The icons on the left side indicate when steps were repeated.}
 \label{fig:procedure}
\end{figure}

The entire evaluation took place in three adjacent rooms (office, body scanner, laboratory) of the University of Würzburg and averaged \SI{117}{\minute}. The procedure is illustrated in Figure \ref{fig:procedure}.

\subsubsection{Opening Phase}
First, participants were informed about the local COVID-19 regulations, received information about the experiment and the body scans, gave their consent, and generated two personal pseudonymization codes used to store the experimental data and the generated avatars separately. Then, the main experimenter answered potential questions and measured the participant's body height without shoes as required for the body scan.
  
\subsubsection{Body Scan Phase}
An auxiliary experimenter performed the body scan in normal clothes without any accessories. Afterwards, the main experimenter measured the interpupillary distance (IPD), body weight, and the participants' waist and hip circumference, and conducted the interview about the scan process. The duration of the interview averaged \SI{4}{\minute}. All interviews were recorded by a Tascam DR-05 voice recorder.

\subsubsection{VR Exposure Phase}
Prior to the VR exposure, participants answered demographic questions and the SSQ as pre-questionnaires on a dedicated questionnaire computer. Then, an auxiliary experimenter demonstrated the participants how to fit the equipment, adjusted the HMD's IPD, and controlled that all equipment was correctly attached. After finishing the fitting, a pre-programmed experimental procedure was started, and participants were transferred to the preparation environment. For all virtual transitions during the VR exposure, the display was blacked-out for a short moment. All instructions were displayed on an instruction panel and additionally played as pre-recorded voice instructions. As the first preparation step, the participants had to undergo a short reading test to ensure the view was sufficient. Then, they performed the embodiment calibration in T-pose and judged its workload. During the whole VR exposure, participants had to answer questions and measurements verbally. Although interaction between the experimenter and the user may cause small breaks in presence \citep{putze2020vrquestions}, we considered this approach as part of the evaluation, since interaction between patient and therapist would also likely occur in clinical settings and advanced in virtuo interaction to answer questionnaires might be difficult for novice users.

After the preparation finished, participants were transferred to the exposition environment. There, they performed five movement tasks in front of a virtual mirror while being instructed to alternatingly look at the mirror and directly on their body to induce the feeling of embodiment. Movement tasks were adapted from related work \citep{wolf2020bodyperception} and had to be performed for \SI{20}{\second}. The first PET followed. Participants estimated the modified body weight of their avatar nine times. Between the estimations, the display was blacked-out briefly to cover the weight changes. In the next phase, participants conducted AMTs nine times for each body weight modification method in a counterbalanced manner. For all body weight estimation tasks, no feedback regarding the estimation error was provided to the participants. 
The second PET concluded the phase. After each AMT (see Figure \ref{fig:procedure}), participants were asked to judge workload, agency, VBO, and VBA in virtuo. The whole VR exposure took \SI{36}{\minute} on average. After the VR exposure, participants answered IPQ, SSQ, UVI, and UX questions again on the dedicated questionnaire computer.

\subsubsection{Closing and Debriefing Phase}
The questionnaires were followed by the second interview about the VR exposure that lasted on average \SI{11}{\minute}. At the end of the session, the main experimenter thanked the participants and granted them credits for their participation. As part of the debriefing process after the session, the interviews were first transcribed and then by two researchers summarized and clustered the answers into categories. 
\section{Results}
In this section, we report the results of our evaluation separated into (1) the body scan experience, (2) the UX of the VR exposure including the different modification methods, and (3) their impact on body image-related measures. The statistical analysis of our results was partially performed using the software R for statistical computing \citep{r2020software} and partially using SPSS version 26.0.0.0 \citep{ibm2020spss}.

\subsection{Body Scan Experience}
\subsubsection{Feedback on the Body Scanning Process}
When asked whether the body scan procedure matched their idea of a body scan, four participants expected a different amount or arrangement of cameras, three participants expected a different scan process (e.g., one camera moving around the body, a laser measuring the body shape, or cameras only in the front), and one participant claimed to have no previous expectations on the body scan process. The other participants stated they already knew the body scan procedure from former experiments and did not remember expectations.

Most of the participants perceived the scan process as simple and clear. Only one participant stated not knowing what was happening between two scans.
The experience during the scan process differed from \enquote{straightforward} and \enquote{easy} ($n=4$) over \enquote{interesting} or \enquote{cool} ($n=4$) to being \enquote{something to getting used to} or making one \enquote{feel observed} ($n=4$). 

All participants stated positively they would do a body scan again. While most of them did not have suggestions for improvement ($n=8$). One suggested that the experimenter should be visible during the whole scan process to increase a feeling of safety. Others pointed out that a reduced number of cameras would ease the feeling of being watched and that the stiff posture during the scan felt kind of uneasy after some time.

\subsubsection{Feedback on the Body Measurements}
When evaluating the assessment of body measures, most participants claimed to perceive it as neutral or similar to being measured during a doctor's appointment ($n=8$). Some others pointed out they would not expect it in a \enquote{normal} lab study but did not perceive it as awkward ($n=3$). One participant stated to perceive the measuring of their weight as very private and therefore uncomfortable. 

\subsection{VR Experience}
Since there was no comparison condition to the overall quantitative scores of the VR experience, we report the data, which were mainly collected on validated and comparable scales, descriptively. For measures captured multiple times during the experience, we calculated the mean value of all data points. The descriptive results of the VR exposure experience are summarized in Table \ref{tab:descriptive-vr}.

\begin{table}[!htb]
  \caption{The table shows the descriptive values for our captured measurements concerning the VR experience. Detailed information regarding the measurements can be found in Section \ref{sec:measures}.}
  \label{tab:descriptive-vr}
  \centering
  \begin{tabu} to \textwidth {X[1,l,m]X[1.3,l,m]X[0.8,l,m]X[0.7,c]}
    \toprule
    Measure & Variable & [Range] & $M$ $(SD)$          \\ 
    \midrule
    Presence  & IPQ General presence & [0-6] & $4.58$ $(0.90)$ \\
     & IPQ Spatial presence & [0-6] & $4.38$ $(0.95)$ \\
     & IPQ Involvement & [0-6] & $3.75$ $(0.89)$ \\
     & IPQ Realism & [0-6] & $3.22$ $(1.2)$ \\
    Embodiment & VBO score & [0-10] & $5.49$ $(2.33)$ \\
     & Agency score & [0-10] & $7.22$ $(1.94)$ \\
    Simulator sickness & Pre-SSQ & [0-235.62] & $26.8$ $(23.7)$ \\
     & Post-SSQ & [0-235.62] & $43.01$ $(39.21)$ \\
    Avatar perception & UVI Humanness  & [1-7] & $4.03$ $(1.10)$ \\
     & UVI Eeriness  & [1-7] & $4.06$ $(0.95)$ \\
     & UVI Spine-tingling & [1-7] & $3.88$ $(0.88)$ \\
     & UVI Attractiveness & [1-7] & $4.25$ $(0.87)$ \\
    Calibration workload & SEA score & [0-220] & $20.83$ $(16.35)$  \\
    Calibration time & Time in \si{\second} & & $96.79$ $(50.29)$  \\
   \bottomrule
\end{tabu}
\end{table}

To evaluate the possible occurrence of simulator sickness, we compared SSQ pre- and post-measurements using a two-tailed Wilcoxon signed-rank test since the pre-requirements for parametric testing were not met. The ratings did not differ significantly between measurements, $Z = 1.14, p = .254$. Additionally, the observed increase in SSQ scores of $16.21$ was below the indication threshold for simulator sickness of $20$ points.

\subsubsection{Feedback on Embodiment and Avatar Perception}
When asked about their feelings towards their personalized avatar, two participants used \enquote{neutral} or \enquote{okay} to describe their experience, and another four participants described it as \enquote{cool}, \enquote{interesting}, or \enquote{pleasant}. The remaining six participants described the experience as less positive, using words like \enquote{strange} and \enquote{irritating}. While one of the former emphasized the quality of the embodiment compared to other studies, three of the latter criticized the embodiment, especially concerning the lack of facial expression, eye movements and hand gestures. One pointed out that their \enquote{hands hold these controllers but the avatar does not}. The participants who found the experience rather irritating emphasized a lack of similarity in their avatar's appearance.

The question of whether the avatar's appearance met the participants' expectations also received mixed responses. While one participant found it overall disproportional, six participants stated that the look of their avatar rather met their expectations. The remaining participants indicated that although the avatar's body looked as expected, they did not associate its face with themselves.

\subsubsection{Comparison of the Body Weight Modification Methods}
For comparing the three AMT conditions (gesture, joystick, and objects), we calculated a one-way repeated-measures ANOVA for each listed measurement (see Table \ref{tab:descriptive-interaction-methods}) except  modification times, where we calculated a Friedman test, and preference rankings, which are presented descriptively only. Test results showed significant differences between conditions only for workload. Two-tailed paired-sample post-hoc t-tests revealed significant differences in the SEA score between body weight modifications with gesture and joystick, $t(11) = 2.74, p = .019$, gesture and objects, $t(11) = 2.8, p = .017$, and joystick and objects, $t(11) = 4.86, p = .001$. Thus, the workload was considered to be highest when modifying body weight via objects and lowest when using the joystick.

\begin{table}[!htb]
  \caption{The table shows all descriptive values of the measures related to the comparison between our proposed body weight modification methods including p-values when calculated. Asterisks indicate significant $p$-values. Post-hoc tests for significant differences can be found in the corresponding text.}
  \label{tab:descriptive-interaction-methods}
  \centering
  \begin{tabu} to \textwidth {X[1.0,l,m]X[0.5,l,m]X[0.5,l,m]X[0.75,c]X[0.75,c]X[0.75,c]X[0.5,c]}
    \toprule
     & & & Gestures & Joystick & Objects & \\ 
    \cmidrule(l){4-4} \cmidrule(l){5-5} \cmidrule(l){6-6}
    Measure & Variable & [Range]  & $M$ $(SD)$  & $M$ $(SD)$ & $M$ $(SD)$ & $p$ \\ 
    \midrule 
    Embodiment & VBO & [0-10]   & $5.75$ $(2.63)$  & $5.08$ $(2.68)$ & $5.38$ $(2.5)$ & $.300$ \\
               & Agency & [0-10]  & $7.25$ $(2.61)$  & $7.16$ $(2.29)$ & $7.33$ $(2.02)$ & $.915$ \\
    Modification time  & Time in \si{\second} & & $23.19$ $(2.94)$  & $24.53$ $(10.37)$ & $27.82$ $(7.72)$ & $.197$ \\
    Workload & SEA & [0-220] & $41.25$ $(27.97)$  & $20.75$ $(13.37)$ & $65.33$ $(33.06)$ & $<.001^\ast$ \\
        & Ranking & [1-4] & $1.91$ & $2.73$ & $3.45$ & -- \\
    Task Difficulty & Ranking & [1-4] & $3$ & $1.81$ & $3.36$ & -- \\
    Vividness & Ranking & [1-4] & $3.09$ & $3.09$ & $2.27$ & -- \\
    Contentment & Ranking & [1-4] & $3.36$ & $3.45$ & $1.91$ & -- \\
    Overall preference & Ranking & [1-4] & $3.27$ & $3.45$ & $2.09$ & -- \\
    Body awareness  & VBA & [0-10] & $6.58$ $(1.98)$  & $7.08$ $(1.93)$ & $6.67$ $(2.06)$ & $.053$ \\
  \bottomrule
\end{tabu}
\end{table}

\subsubsection{Feedback on the Body Weight Modification Methods} 
When asked to explain their preference for an interaction method, most of the participants who preferred joystick ($n=8$) stated that it felt most controllable and least complicated. One participant additionally preferred the continuity of joystick-based interaction compared to the necessity of repetition in the gesture-based interaction. The participants who had preferred the gesture-based interaction ($n=4$) stated they found it most intuitive, flexible, and direct. They reasoned that controlling the speed of modification by extent and speed of arm movements increased usability. None of the participants preferred modification via the objects.
xx
\subsection{Body Image}
In the following, we present the impact of our VR exposure on the body image-related measures of body awareness and body weight estimation as well as the related qualitative feedback. 

\subsubsection{Comparison of Body Awareness Between Body Weight Modification Methods} 
We calculated a one-way repeated-measures ANOVA to compare the body awareness (VBA) during the three AMT conditions (gesture, joystick, and objects). As shown in Table \ref{tab:descriptive-interaction-methods}, VBA ratings differed tendentially between the three AMT conditions, with higher joystick ratings than the other conditions.

\subsubsection{Feedback on the Intensity of Body Awareness}
Seven participants stated they felt in contact with their physical body during the experience, while the other five stated they had lost contact to their body at least once. The latter stated, for example, they focused mainly on the task and the avatar. 
Others felt that their bodily awareness \enquote{got a bit lost} or that the situation and virtual surroundings made them forget reality, including their real body. On the other hand, three participants who stated being aware of their body during the experiment reasoned the embodiment as a main cause. One of them stated that \enquote{once before re-calibration, my avatar's foot was kind of crooked, that's when I paid attention to my real body. I made sure my knee was straight}. The other one focused on the avatar weight and claimed that \enquote{I was still aware of my body, but it was very strange because I was looking at a different mirror image, and sometimes, I felt much heavier when the weight of the avatar was lower than my actual weight}. Another reason why participants were aware of their bodily sensations was the physical contact with the floor or the proprioception during movements, which reminded them of their presence in the physical room ($n=2$).

\subsubsection{Feedback on the Affect Towards the Body}
Eight of the participants stated that their feelings towards their bodies had changed during the experience. These changes concerned either their general awareness ($n=3$), their experienced body size ($n=2$), or their satisfaction with their body ($n=3$). The two participants stating a change in their experienced body size had either felt thicker or thinner in contrast to their avatar during the experience or felt thinner after the experience. Two of the participants whose bodily satisfaction changed stated an increased body satisfaction or increased motivation to care for their bodily interests. In contrast, one participant reported increased dissatisfaction towards their physical body after the experience.

\subsubsection{Feedback on the Perception of Body Weight Changes} 
Concerning the changes in the avatar's body weight, the participants equally rated them as \enquote{interesting} ($n=6$) or \enquote{weird} ($n=6$). Two participants especially pointed out that it was interesting to compare the avatar's body shape to their own former body, as they had lost or gained weight in the past. One stated \enquote{when I started my studies five years ago, I was $\SI{20}{\kg}$ lighter than now, and it was kind of interesting to compare the avatar's look to the memories of my old body shape. It gave me a little perspective on how I want to look}. Four of the other participants liked the idea of seeing how they could look if they changed their eating/exercise behavior. Especially the modification towards a lower weight was perceived as threatening by some of the participants ($n=3$), as they thought it looked a bit unhealthy. To enhance the modification, two participants suggested more individual and fine-grained possibilities to manipulate only body parts instead of the body as a whole, for example, by including \enquote{two fixed points on the virtual body, one in the middle of the body and one at the shoulder area, to adjust the weight in these areas more exactly}.

\subsubsection{Comparison of Body Weight Estimations between Body Weight Modification Methods}
For comparing the performance in body weight estimations between the AMT, we calculated a one-way repeated-measures ANOVA for $\overline{M}$-values, the percentage body weight misestimation, and a Friedman test for $\overline{A}$-values, the absolute percentage body weight misestimation. The tests revealed that the three interaction methods did not differ significantly, neither in $\overline{M}$ nor in $\overline{A}$, as summarized in Table \ref{tab:descriptive-body-weight-perception}. 

\begin{table}[!htb]
  \caption{The table summarizes the body weight estimation performance (average misestimation $\overline{M}$ and absolute average of misestimation $\overline{A}$)  of the comparison between our proposed modification methods.}
  \label{tab:descriptive-body-weight-perception}
  \centering
  \begin{tabu} to 0.55\textwidth {X[0.7,c]X[1,c]X[1,c]X[1,c]X[0.3,c]}
    \toprule
    & Gestures & Joystick & Objects & \\ \cmidrule(l){2-2} \cmidrule(l){3-3} \cmidrule(l){4-4}
    & $M$ $(SD)$ & $M$ $(SD)$ & $M$ $(SD)$ & $p$  \\ 
    \midrule
    $\overline{M}$ in \si{\percent} & $3.44$ $(9)$ & $3.44$ $(8.9)$ & $2.41$ $(8.05)$ & $.529$ \\
    $\overline{A}$ in \si{\percent} & $8.92$ $(4.58)$ & $8.46$ $(5.10)$ & $8.36$ $(3.66)$ & $.780$ \\
   \bottomrule
\end{tabu}
\end{table}

\subsubsection{Comparison of Body Weight Estimations Between Estimation Methods}
We compared AMT and PET using two-tailed paired-samples t-tests for $\overline{M}$-values and two-tailed Wilcoxon signed-rank tests for $\overline{A}$-values. For $\overline{M}$, we showed that participants misestimated the body weight significantly less using the PET. For $\overline{A}$, we could observe a clear tendency towards lower misestimations for PET. Results are summarized in Table \ref{tab:descriptive-modification-task}.

\begin{table}[!htb]
  \caption{The table summarizes the body weight estimation performance (average misestimation $\overline{M}$ and absolute average of misestimation $\overline{A}$)  of the comparison between passive estimation task (PET) and active modification tasks (AMT). Asterisks indicate significant $p$-values.}
  \label{tab:descriptive-modification-task}
  \centering
  \begin{tabu} to 0.5\textwidth {X[0.7,l,m]X[1,c]X[1,c]X[0.6,c]}
    \toprule
    & PET & AMT & \\ 
    \cmidrule(l){2-2} \cmidrule(l){3-3} 
    & $M$ $(SD)$  & $M$ $(SD)$ & $p$ \\ 
    \midrule
    $\overline{M}$ in \si{\percent} & $1.46$ $(8.4)$ & $3.1$ $(8.4)$ & $.031^\ast$  \\
    $\overline{A}$ in \si{\percent} & $7.74$ $(4.43)$ & $8.59$ $(4.14)$ & $.060$  \\
  \bottomrule
\end{tabu}
\end{table}

\begin{figure}[!htb]
	\centering
	\caption{The figure shows the body weight misestimations $M$ (left) and absolute body weight misestimations $A$ (right) in relation to the performed body weight modifications for PET and AMT.}
	\label{fig:bodyweight-modification-trials-regression}
	\includegraphics[width=0.49\textwidth,trim={0mm 0mm 0mm 0mm},clip]{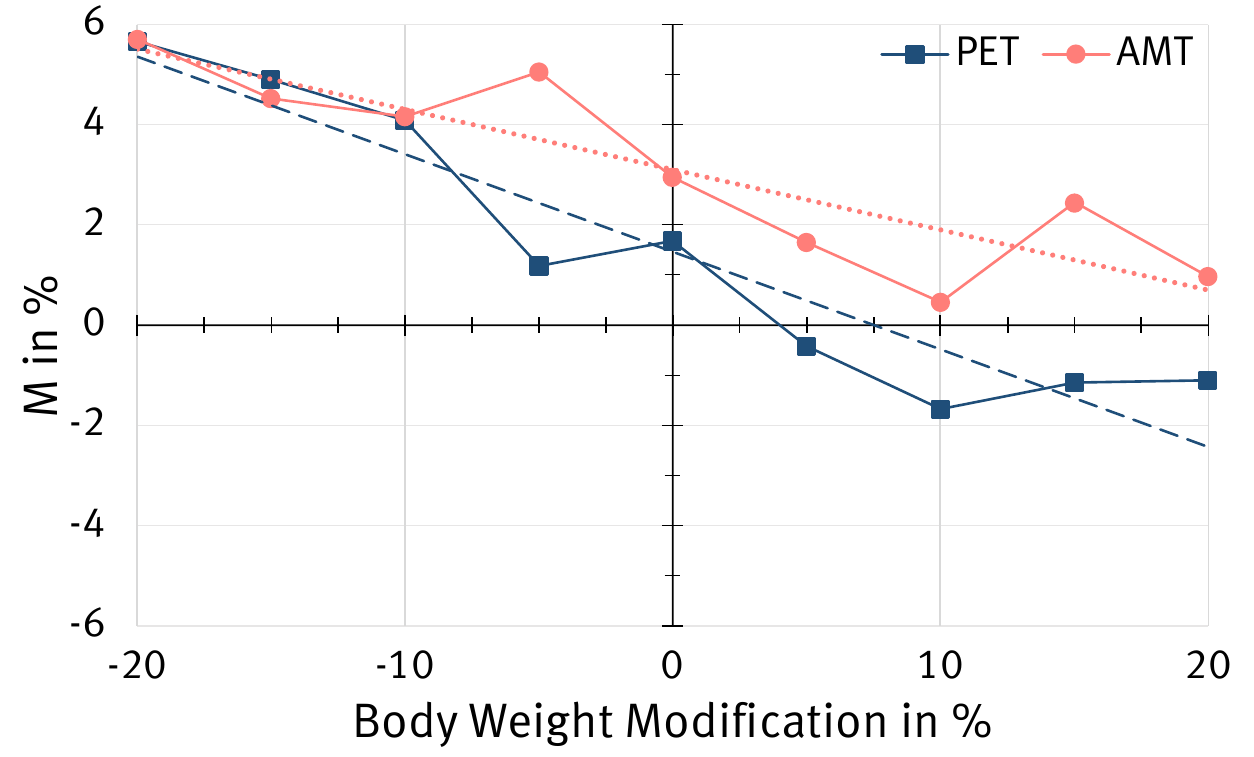}
	\includegraphics[width=0.49\textwidth,trim={0mm 0mm 0mm 0mm},clip]{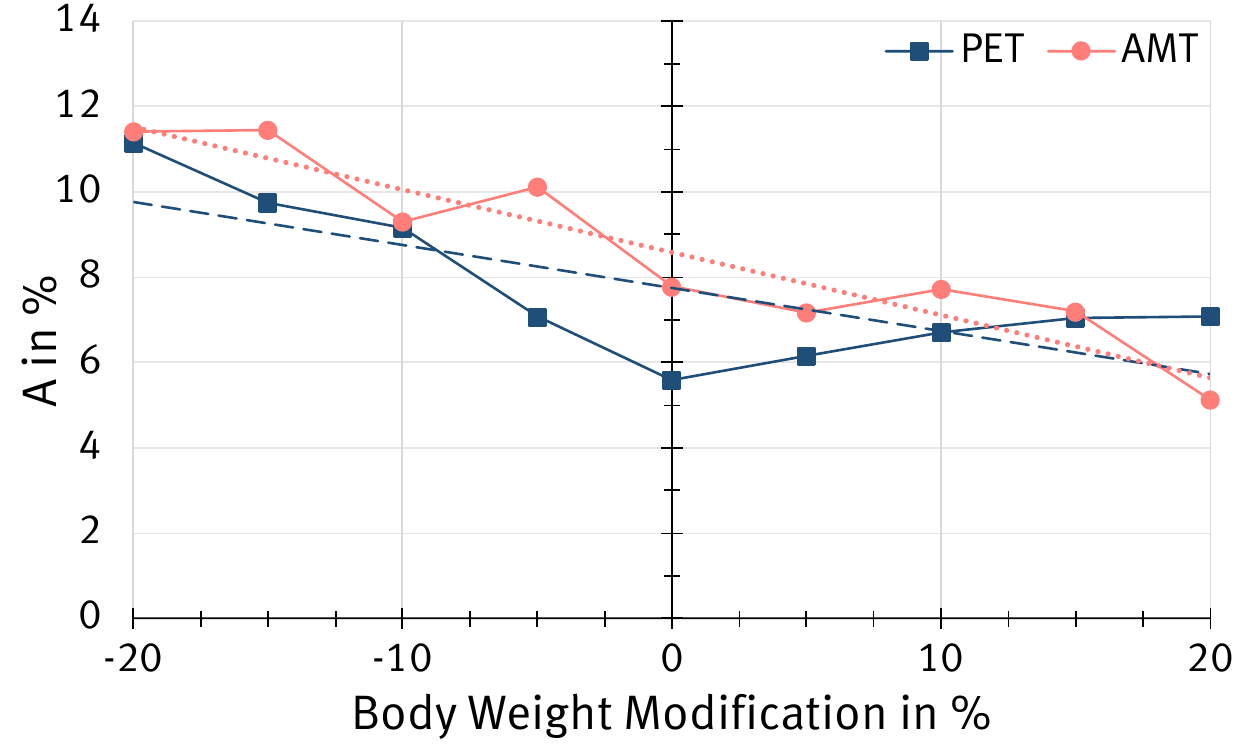}
\end{figure}

We further analyzed the results of AMT and PET concerning the modification levels ($\pm 20\%$ in $5\%$ steps) using linear regression. Our data violated pre-requirements for linear regression in terms of homoskedasticity and normality. Therefore, we calculated each linear regression using parameter estimations with robust standard errors (HC4) as recommended by \citet{hayes2007heteroskedasticity}. Figure \ref{fig:bodyweight-modification-trials-regression} shows the body weight misestimations $M$ (left) and the absolute body weight misestimations $A$ (right) for PET and AMT in relation to the modification levels.

For $M$, the results showed a significant regression equation for PET, $F(1,106) = 7.88, p = .006$, with a \textit{$R^2$}~of~.069. The prediction followed the equation $\textrm{M} = -0.194 \cdot \textrm{Body Weight Modification in \%}+1.462$. The modification level did significantly impact on body weight misestimations $M$, $\textit{t}(106)~=~-5.11, \textit{p}~=.013$. For AMT, we found no significant prediction of the modification level on the body weight misestimations $M$, $F(1,106) = 3.05, p = .084$, having a \textit{$R^2$}~of~.028. The found prediction followed the equation $\textrm{M} = -0.120 \cdot \textrm{Body Weight Modification in \%}+3.099$. In consequence, the modification level did not significantly impact on body weight misestimations $M$, $\textit{t}(106)~=~-3.46, \textit{p}~=.094$. 

For $A$, the results showed a significant regression equation for PET, $F(1,106) = 5.27, p = .024$, with a \textit{$R^2$}~of~.047. The prediction followed the equation $\textrm{A} = -0.101 \cdot \textrm{Body Weight Modification in \%}+7.743$. The modification level did significantly impact on body weight misestimations $A$, $\textit{t}(106)~=~-2.09, \textit{p}~=.039$. For AMT, we found a significant prediction of the modification level on the body weight misestimations $A$, $F(1,106) = 15.7, p < .001$, with a \textit{$R^2$}~of~.129. The found prediction followed the equation $\textrm{M} = -0.147 \cdot \textrm{Body Weight Modification in \%}+8.585$. The modification level did significantly impact on body weight misestimations $A$, $\textit{t}(106)~=~-17.9, \textit{p}~<.001$.

In addition to the linear regressions, we averaged negative and positive modifications for both measurements to analyze difference between the modification directions. Test results for $M$-values showed that body weight misestimations differed significantly between positive ($M=-1.09$, $SD=7.44$) and negative($M=3.96$, $SD=11.13$) modifications for PET, $t(11) = 2.27, p = .044$, but not between positive ($M=1.38$, $SD=7.45$) and negative ($M=4.86$, $SD=10.57$) modifications for AMT, $t(11) = 1.63, p = .131$. For $A$-values, we found no significant differences between positive ($M=6.74$, $SD=3.5$) and negative ($M=9.28$, $SD=7.2$) modifications for PET, $Z = 1.26, p = .209$, but found significant differences between positive ($M=6.8$, $SD=4.4$) and negative ($M=10.57$, $SD=4.9$) modifications for AMT, $Z = 2.59, p = .010$.

\subsubsection{Feedback on the Body Weight Estimation Difficulty} 
Regardless of the estimation method, estimating the body weight of the avatar was found to be difficult ($n=8$). Only three participants stated they found it relatively easy or only medium-difficult to estimate the body weight. The main reason why participants rated the task as difficult was the high number of repetitions ($n~=~2$) or a reduced perceptibility of their physical body, both leading to a \enquote{loss of perspective}. Additionally, one participant stated that the task difficulty depended on the distance of the avatar's weight to their own.
\section{Discussion}
In the present paper, we introduced a prototype of an interactive VR-based system that aims to support body image interventions based on embodied, modulatable, and personalized avatars in future clinically relevant settings. We evaluated the system regarding (1) the body scan experience, (2) the UX of the VR exposure including body weight modification interaction methods, and (3) the impact on two body image-related measures body awareness and body weight misestimation. In the following, we summarize and discuss the results of our evaluation to ultimately derive guidelines supporting the design of systems for body image interventions. The guidelines are based on conclusions of the qualitative and quantitative results accomplished by the researchers' observations and participants' comments during the evaluation. While these may overlap with existing best practices or established VR guidelines, we believe it is elementary to summarize them for the given context and to highlight their importance.

\subsection{Body Scan Experience}
Overall, the scan process was mainly rated as simple and interesting, although it took place in a separate room with great technical effort. Participants stated a high acceptance and willingness to be scanned again. In addition, the scan and the associated body measurements were seen as something that one would do in a clinical setting, and that does not trigger unpleasant reservations. This assessment strengthens the idea of using body scans in a clinical context.

Nevertheless, two main criticisms of the scanning process were the feelings of being watched and being left alone. The large number of visible cameras mainly caused the first while both can be attributed to the arrangement of the cameras surrounding the person in all directions. Curtains around the scanner also supported the feeling of being left alone during the scan process. In particular for our target group and the intended clinical usage, amendments seem necessary. Options to reduce the negative feelings could be a change in the arrangement of cameras, e.g., opening the space by placing them only on one side or reducing the number of cameras to a minimum as proposed by \citet{wenninger2020smartphone} and supported by the results of \citet{bartl2021lowcost}. In addition, we suggest a constant dialogue about and during the process to counteract the feeling of being alone.

\noindent \textbf{Guidelines for Body Scanning}
\begin{itemize}
    \item Users should receive thorough information and instruction in advance about the body scan procedure to provide clarity and transparency.
    \item Body scans should be performed unobtrusively to protect privacy and avoid the feeling of being watched.
    \item The number and arrangement of cameras should be planned carefully to avoid the feeling of being watched.
    \item The number of people involved in the body scan should be minimized to increase privacy, and personal contact should be maximized to increase safety.
    \item Body-related measurements should be performed professionally while maintaining privacy.
\end{itemize}

\subsection{User Experience of VR Experience}
The feedback regarding preparation and calibration was consistently positive, confirming the decision for our approach. This is empirically supported by the low measured calibration times requiring only a short time holding T-pose, and the low workload ratings during the calibration process. Nevertheless, there are further possibilities to reduce the effort for calibration and invasiveness, for example, by using completely markerless body tracking solutions \citep[cf.,][]{wolf2022holographic}.

Regarding VR-specific measures, participants rated their perceived feeling of presence on an acceptable level \citep[cf.,][]{buttussi2018presence,wolf2020bodyperception}, with lower ratings on involvement and realism. A reason for the lower observed involvement score could be the constant interaction with the experimenter during the tasks (e.g., confirming body weight estimations, rating experiences). Possible implausible content (e.g., body weight modification by interaction) could have impacted negatively on realism. Continuous communication between therapist and patient during weight modifications might be a crucial element in clinical settings. Therefore, further research on the role of presence (and its sub-dimensions) in VR body image interventions seems required, as the latest reviews did not address this topic \citep{turbyne2021review,riva2019virtual,horne2020review}.

Surprisingly, although participants rated their feeling of virtual body ownership descriptively  higher compared to non-personalized avatars \citep{waltemate2018impact,wolf2020bodyperception}, their ratings were lower than in prior work using personalized and photorealistic avatars \citep{waltemate2018impact}. A reason for the noticed differences could be the particularly body-related nature of our task. Avatars created via body scans have a very high resemblance to the individual but still do not provide a perfect visual replica. In a task highly focusing on body perception, even minor inaccuracies may become noticeable, and participants might focus on these, experiencing a diminished feeling of virtual body ownership. Another factor could be the performed body weight modification, as real-time changes in the avatar's body shape lead to a reduced concordance between real and virtual bodies.

The ratings and especially the qualitative statements on avatar perception reveal similar results, as some of the participants stated their avatar to be uncanny or not fully recognizable as themselves. This raises doubts about the degree of personalization of avatars and whether the creation of highly photorealistic textures is currently necessary (and feasible). Tools such as Virtual Caliper \citep{pujades2019caliper} can create in shape personalized avatars using only VR equipment. In conjunction with generic avatar generators, such as Meta Human \citep{unreal2021metahuman}, highly realistic avatars with personalized body shapes could be created. They wouldn't resemble the person perfectly. However, this lack of resemblance could make them less uncanny while remaining a still better quality in general. Additionally, a personalization in body shape would be sufficient for simulating body weight changes. One counter-argument is provided by \citet{thaler2018bse}, who clearly state that the body weight perception of avatars having personalized textures differs from generic ones. More research in this direction seems required.

\noindent \textbf{Guidelines for VR Design}
\begin{itemize}
    \item The physical and mental effort for system calibration should be kept as low as possible.
    \item The animations of embodied avatars should be as authentic as possible and include facial expressions, eye movements, and hand gestures to increase realism and reduce eeriness.
    \item When using physical controllers, virtual controller representations should be displayed in VR and controlled by the avatar.
    \item When using personalized avatars, body shape and texture should aim for the highest possible conformity with the user to reduce uncanniness.
\end{itemize}

\subsubsection{User Experience of Body Weight Modification}
When comparing the subjective rankings of the three modification methods, it becomes apparent that the interaction via virtual objects was the least preferred. It was rated as more demanding and difficult, and less vivid, resulting in lower contentment and overall preference than the other two modification methods. Modification via joystick and gestures were rated rather similarly with a slight preference towards the joystick interaction. The in virtuo ratings of workload match these rankings. While joystick was rated quantitatively most positively, the qualitative analysis shows arguments in favor of gesture interaction, especially in terms of vividness and intuitivity. No impact has been noticed on the feeling of embodiment or performance in body weight estimation, which is particularly important in our context.

Regardless of the interaction method, the lack of body weight modification in relation to different body parts (e.g., abdomen, hips, thighs) and in relation to the composition of the body tissue (e.g., fat or muscle mass) was mentioned. The use of advanced body modification methods, such as those presented by \citet{maalin2020beyond} or \citet{pujades2019caliper} could allow for body weight modifications that go beyond using only BMI as a single parameter modifying the whole body's weight. However, having more complex body weight modification methods would also increase the difficulty of user interaction.

\noindent \textbf{Guidelines for Body Weight Modifications}
\begin{itemize}
    \item Body weight modifications severely differing from the user's BMI or reaching unrealistic ranges should be avoided to reduce alienation.
    \item Body weight modifications should allow changing the body weight independently on different body parts considering different body tissue compositions.
    \item Body weight modifications performed directly via a hardware input device or body gestures should be preferred over virtual objects or buttons.
\end{itemize}

\subsection{Body Image-Related Outcomes}
The reported effects of the VR exposure on body awareness and affect towards the body were very individual, with participants reporting either a loss or an increase of body awareness and either an affirmation of their positive body image or an increase of negative feelings towards their body. Future work with an increased sample size is necessary to investigate whether these individual differences are related to people's overall body awareness, as proposed by \citet{filippetti2017heartfelt}, and to a negative or positive body image. These insights will be crucial to determine what effects can be expected for a target group with low body awareness or negative body image. The comparison of body awareness between the three modification methods indicates a higher body awareness in joystick interaction, followed by gestures and objects. Compared to the results on workload, this suggests that task load might harm body awareness in VR. 

In contrast to body awareness, body weight estimations did not differ between body weight modification methods. However, when comparing the accuracy of the type of estimations task, PET provided more accurate estimates than AMT. While estimating a person's weight based on their appearance is not an everyday task, it is surely more common than actively modifying a (virtual) body to a certain body weight. Thus, the difference might originate in the relative novelty of active modification compared to passive estimation. Another reason could be the different phrasing of the task instructions, which has been shown to have the ability to influence body weight estimation \citep{piryankova2014owning}. For both PET and AMT, the accuracy of the body weight estimation depended on the target weight, or in other words, on the deviation between the own real weight and the virtually presented body weight. This effect has been observed priorly for VR body weight estimation tasks \citep{wolf2020bodyperception,thaler2018bse} and is in line with the so-called \emph{Contraction bias} as described by \citet{cornelissen2016visualbias,cornelissen2015bmiinfluence}. It states that body weight estimates are most accurate around an estimator-dependent reference template (of a body) and decrease with increasing BMI difference from this reference. Thereby, bodies heavier than the reference tend to be underestimated, while lighter ones tend to be overestimated. Results on absolute body weight estimations show that although the average misestimations were comparatively low, they are subject to high deviations and uncertainties, which also has been observed priorly \citep{thaler2018bse}. The reasons for this probably lie in the nature of the task, since estimating body weight seems generally challenging, and body image disturbances are ubiquitous even in the healthy population \citep{longo2017distorted}. Qualitative feedback confirms the task difficulty. When further analyzing the absolute body weight estimations, it is particularly noticeable that they seem to be easier and more accurate for increased than for reduced body weight. This is rather unexpected since \emph{Weber's Law} suggests that differences in body weight become harder to detect when body weight increases \citep{cornelissen2016visualbias}. A possible reason for the high uncertainties in the absolute body weight estimations and the contradiction to Weber's law could be the perspective on the avatar offered by the virtual mirror, which mainly shows the front side of the body \citep{cornelissen2018viewbias}. More research on this topic seems required.

\noindent \textbf{Guidelines for Body Weight Estimations}
\begin{itemize}
    \item Body weight estimations capturing the current perception of the real body in VR should be performed at the beginning of an intervention, as the perceptibility of the real body might decrease over time.
    \item When performing body weight estimations, care should be taken to present the respective body equally from multiple perspectives.
    \item When analyzing body weight misestimations based on avatars, determining the average accuracy of the misestimations with healthy individuals helps avoid strong influences of the system properties.
\end{itemize}

\subsection{Future Research Directions}
The results of our work raise new research questions for future work. 
First, the high necessity of communication between therapist and user, potentially leading to breaks in presence, raises the question of the general impact of presence in body image interventions. This is also interesting when it comes to augmented reality, as already recognized by \citet{wolf2022holographic}.

Second, the observed ratings in body ownership despite using photorealistic, personalized avatars and the feedback on avatar perception leads to the question of how photorealism and personalization should be applied to body image interventions. Future work should explore whether avatars that are less personalized in texture  are sufficient for our purpose as they might raise less uncanniness. 

Third, the severe individual differences in the report of body awareness and affect towards the body raise the question, of which individual characteristics might predict the effects of a VR-based intervention on both variables. Further, future work should explore the effects of task difficulty on body awareness and whether it mediates the effect of a VR-based intervention on body image.  


Finally, future work should further address the difference between active body weight modification and passive body weight estimation we found in this study. It remains unclear which underlying processes lead to differences between the two tasks and whether they impact differently on body image. Similar counts for the observed differences in body weight misestimations for avatars with decreased or increased body weight.

\subsection{Limitations}
Although our sample included slightly overweight participants, the current design and development phase was limited to students without a diagnosed body image disturbance and predominantly with a BMI in a healthy range. The clinical applicability to our target group, which is already in preparation as part of our ViTraS research project \citep{dollinger2019vitras}, is consequently the next step after the here presented design and UX optimization phase. Also, the limits of body weight modification used in this study were selected arbitrarily. Some participants felt uneasy when their avatar's body weight was modified, especially in the extreme ranges. 
Overall, the design and development phase would benefit from a larger test user group tailored to the final target group. However, this is not an easy endeavor since it blurs the separation between the user-centered design and development phases and the clinical application. Hence, it requires closer integration and supervision by therapeutically trained professionals and experts in obesity treatment. Ultimately, this integration would be necessary throughout all steps of the complete user-centered technical developments to safeguard against unwanted effects for all participants during the design and development and UX optimization steps. Notably, two participants of our overall healthy sample already showed some emotional reactions when confronted with their modified virtual self.
xx
\section{Conclusion}
In this work, we have presented and evaluated the prototype of an advanced VR therapy support system that allows users to embody a rapidly generated, personalized, photorealistic avatar and modulate its body weight in real-time. Our system already offers numerous positive features and qualities, especially regarding the execution of body scans and an overall enjoyable VR experience. The guidelines for designing VR body image therapy support systems we derived from our results help facilitate future developments in this field. 

However, more research is needed for a therapeutic application. Possible areas of investigation include the implementation of photorealism, which may need to be revisited when working on body image. More research is also required on the differences between active body weight modification and passive body weight estimation. Finally, investigations with more focus on the target group and the individual characteristics of future users will be necessary, especially concerning body image distortion, body dissatisfaction, and body awareness.

\section*{Conflict of Interest Statement}
The authors declare that the research was conducted in the absence of any commercial or financial relationships that could be construed as a potential conflict of interest.

\section*{Author Contributions}
ND and EW conceptualized large parts of the experimental design, collected the data, performed data analysis, and took the lead in writing the manuscript. EW and DM developed the Unity application including the experimental environment and avatar animation system. MB and SW provided the avatar reconstruction and body weight modification framework. CW and ML conceived the original project idea, discussed the study design, and supervised the project. All authors continuously provided constructive feedback and helped to shape study and the corresponding manuscript.

\section*{Acknowledgments}
We thank Andrea Bartl for her extensive support when preparing and conducting the body scans, Viktor Frohnapfel for contributing his Blender expertise to our virtual environments, Marie Fiedler for proofreading, and Sara Wolf for her support with our illustrations. We also thank Miriam Fößel and Nico Erdmannsdörfer for their help in preparing the interviews for qualitative analysis. In addition, we would like to thank the project partners from the ViTraS research project for their constructive feedback.

\section*{Funding}
This research has been funded by the German Federal Ministry of Education and Research in the project ViTraS (project number 16SV8219).


\section*{Data Availability Statement}
All datasets for this study can be provided upon request.

\bibliographystyle{frontiersinSCNS_ENG_HUMS} 
\bibliography{references}






\newpage
\appendix
\section{Evaluation Interview Questions}
\subsection{Body Scan Experience}
Questions about the body scan:
\begin{enumerate}
\item What expectations did you have before the participation about how a body scan would be?
\item Was it clear at any point during the body scan what you had to do?
    \begin{enumerate}
    \item If not: At what point were there ambiguities and how did they arise?
    \end{enumerate}
\item How did you feel about the scanning process? 
\begin{enumerate}
    \item What was the reason for it?
    \item Were those feelings more pleasant or unpleasant?
    \end{enumerate}
\item Would you decide to get your body scanned again?
    \begin{enumerate}
    \item If not: What is the reasoning behind it?
    \end{enumerate}
\item If you could change something about the scanning process, what would it be?
\item Did the gender of the experimenter affect how comfortable/uncomfortable you felt during the scan?
\item Would you have felt differently about the scanning process if you had known the experimenter better?
\end{enumerate}

\noindent Questions about the body measurements:
\begin{enumerate}
\item How did you feel about the body measurements being taken?
    \begin{enumerate}
    \item What was the reason for it?
    \item Were those feelings more pleasant or unpleasant?
    \end{enumerate}
\item If you could change something about the body measurement process, what would it be?
\item Did the gender of the experimenter affect how comfortable/uncomfortable you felt during the body measurements?
\end{enumerate}

\noindent Questions about the body scan and body measurements:
\begin{enumerate}
\item What could the experimenter have done to make you feel more comfortable during the scanning and measurement process?
\end{enumerate}

\subsection{VR Exposure Experience}
\noindent Questions about the interaction with the embodied virtual human (avatar):
\begin{enumerate}
\item How did you feel about the interaction with your personal avatar?
    \begin{enumerate}
    \item What was the reason for it?
    \item Were those feelings more pleasant or unpleasant?
    \end{enumerate}
\item Did the appearance of your avatar meet your expectations?
    \begin{enumerate}
    \item If not: What would you have expected differently?
    \item If not: Was the deviation from your expectation positive or negative?
    \end{enumerate}
\item Did you find it rather easy or rather difficult to estimate the weight of your avatar when it changed without your action?
    \begin{enumerate}
    \item If difficult: What was the reason for it?
    \end{enumerate}
\item Did you find it rather easy or rather difficult to adjust your avatar to the given weight?
    \begin{enumerate}
    \item If difficult: What was the reason for it?
    \end{enumerate}
\item Is there one method of interaction that you would prefer over the others?
    \begin{enumerate}
    \item What was the reason for it?
    \end{enumerate}
\item If you could make something about the interaction with your avatar different, what would it be?
\end{enumerate}

\noindent Questions about the (physical) experience:
\begin{enumerate}
\item How did it feel for you when the appearance of your personal avatar changed?
    \begin{enumerate}
    \item Did it feel different when you actively changed the appearance of your personal avatar?
    \end{enumerate}
\item Were you aware of your physical body while being embodied to your virtual avatar? 
    \begin{enumerate}
    \item If yes: Were there moments when you paid particular attention to your physical body?
    \end{enumerate}
\item Do you had the feeling that interacting with your avatar had an impact on how you felt in your physical body?
    \begin{enumerate}
    \item If yes: In what ways did you feel changed?
    \end{enumerate}
\item Did the interaction with your in body weight changed avatar cause you to perceive or see your own body differently?
    \begin{enumerate}
    \item If yes: What has changed?
    \item Do you take any direct consequences from this experience?
    \end{enumerate}
\item Could you imagine an interaction in which the virtual avatar supports you in experiencing your physical body more consciously?
    \begin{enumerate}
    \item If yes: How would it look like?
    \end{enumerate}
\end{enumerate}

\noindent Questions for the instruction of the tasks:
\begin{enumerate}
\item What were your expectations about how you would receive instruction within the virtual environment?
\item How did you feel that the instructions for the tasks were given verbally and in text form?
    \begin{enumerate}
    \item What was the reason for it?
    \end{enumerate}
\item Did you notice that there was no visual representation in the form of a speaker or something similar for the verbal instructions?
    \begin{enumerate}
    \item If yes: Was it rather pleasant or rather unpleasant?
    \item Where in the virtual environment did you locate the instruction?
    \end{enumerate}
\item If you imagine a visual representation of the instructing voice, how would it look like?
\item Can you imagine sharing the virtual environment with another person while changing the appearance of your virtual avatar?
\end{enumerate}

\noindent Questions about the overall process:
\begin{enumerate}
\item If you had the choice - what would you change about the overall process?
    \begin{enumerate}
    \item Did you notice anything else?
    \end{enumerate}
\end{enumerate}

\end{document}